\documentclass[a4paper,aip,apl,amsmath,amssymb,reprint,superscriptaddress]{revtex4-1}
\usepackage{graphicx}
\usepackage{amsmath}
\usepackage{dcolumn}
\usepackage{bm}
\usepackage{bbold}
\usepackage{color}
\usepackage{amsfonts}
\usepackage{amssymb}
\usepackage{mathrsfs}
\usepackage{tabularx}
\usepackage{braket}
\usepackage{mathtools}
\usepackage{soul}			

\usepackage{caption}
\usepackage{subcaption}

\begin{document} 
\title{Axion Detection with Precision Frequency Metrology}
\author{Maxim Goryachev}
\affiliation{ARC Centre of Excellence for Engineered Quantum Systems, School of Physics, University of Western Australia, 35 Stirling Highway, Crawley WA 6009, Australia}

\author{Ben T. McAllister}
\affiliation{ARC Centre of Excellence for Engineered Quantum Systems, School of Physics, University of Western Australia, 35 Stirling Highway, Crawley WA 6009, Australia}

\author{Michael E. Tobar}
\email{michael.tobar@uwa.edu.au}
\affiliation{ARC Centre of Excellence for Engineered Quantum Systems, School of Physics, University of Western Australia, 35 Stirling Highway, Crawley WA 6009, Australia}

\date{\today}


\begin{abstract}

We investigate a new class of galactic halo axion detection techniques based on precision frequency and phase metrology. Employing equations of axion electrodynamics, it is demonstrated how a dual mode cavity exhibits linear mode-mode coupling mediated by the axion upconversion and axion downconversion processes. The approach demonstrates phase sensitivity with an ability to detect axion phase with respect to externally pumped signals. Axion signal to phase spectral density conversion is calculated for open and closed loop detection schemes. The fundamental limits of the proposed approach come from the precision of frequency and environment control electronics, rather than fundamental thermal fluctuations allowing for table-top experiments approaching state-of-the-art cryogenic axion searches in sensitivity.
Practical realisations are considered, including a TE-TM mode pair in a cylindrical cavity resonator and two orthogonally polarised modes in a Fabry-P{\'e}rot cavity.


\end{abstract}

\maketitle

\section*{Introduction}

Axions are theoretical weakly-interacting sub-eV particles~\cite{wisps} that can be formulated as a primary component of dark matter. With mounting evidence pointing towards lower mass particles~\cite{NatureDM1,NatureDM2}, particles such as axions are becoming increasingly promising dark matter candidates. Axions arise as a result of an elegant solution to the strong CP problem in QCD~\cite{AxionCP}, and are expected to have properties consistent with dark matter~\cite{AxionDM}. Confounding experimental efforts to detect axions is the fact that the axion mass is largely unknown, with only weak bounds from cosmology and theory. Despite this, a number of axion detection experiments are already underway~\cite{ORGAN,ADMX2010,CAPP,CAPPToroid,MADMAX,HAYSTAC}.

In perhaps the most common axion detection technique, it is generally agreed that dark matter axions can be detected via the (inverse) Primakoff effect, a two photon-axion interaction. Nowadays, in a typical (Sikivie) detector\cite{Sikivie:1983aa}, these hypothetical particles interact with a DC magnetic field, or virtual photons, to produce real photons whose frequency corresponds to the mass of axions. This scheme employs one or several tuneable microwave cavities, serving as resonant antennas, with the output coupled to the lowest noise amplifiers so generated photons may be detected with the greatest sensitivity possible. In principle, a similar detector using static electric field is also possible although due to considerable mismatch between electric and magnetic components and relative difficulty in creating extremely strong electric fields in large volumes, such detectors are never realised in practice. The third Sikivie-like axion detection technique is represented by the type of detectors utilising RF or microwave fields instead of static ones: indeed, the Primakoff process works equally well with real photons instead of virtual ones\cite{Melissinos2009,RfSikivie}. In these schemes the existing photons of a given frequency interact with cosmic axions creating additional photons with a different frequency (such that energy is conserved) at a rate related to the number of pre-existing photons and the number of axions. Prior work has considered a single pumped RF cavity mode interacting with axions, followed by detecting a small signal in an orthogonally polarised mode\cite{Melissinos2009,RfSikivie}. In this work we consider more general cases of two mode-axion interactions.

Despite the apparent similarity between the DC and AC detection schemes, they belong to different classes of detectors. Since virtual photons or static fields carry no phase, the traditional Sikivie haloscope detectors (using DC magnetic or electric fields) belong to the class of phase insensitive systems. On the other hand, the AC scheme considered in this work relies on pumping signal(s) carrying relative phase as well as separate phases relative to the axion signal. Thus, the detected signal as well as the overall result would have a footprint of these phases. This fact draws analogies with existing amplifiers\cite{Caves:1982aa} that can be grouped into DC (phase insensitive) amplifiers, where energy is drawn from static power supply, and parametric (phase sensitive) amplifiers, where energy comes from oscillating fields. The second type gives more freedom allowing improved amplification/detection schemes based on quadrature squeezing\cite{Caves:1982aa}. Thus, in this work we expand investigations into a novel class of axion detectors employing the phase sensitive approach. 

\section{Axion Electrodynamics Description}

The Hamiltonian density of the photon-axion system consist of the conventional electromagnetic, axion and interaction parts:
\begin{multline}
\begin{aligned}
	\label{C001CO}
	\displaystyle  \mathcal{H} = \mathcal{H}_{\text{EM}} + \mathcal{H}_{a} + \mathcal{H}_{\text{int}}.
\end{aligned}
\end{multline}
The free electromagnetic Hamiltonian density is usually represented using the vectors of electric and magnetic fields,  $\mathbf{E}$ and $\mathbf{B}$ respectively, or a vector potential and its conjugate momentum, $\mathbf{A}$ and $\mathbf{\Pi}$: 
\begin{multline}
\begin{aligned}
	\label{C002CO}
	\displaystyle  \mathcal{H}_{\text{EM}} = \frac{\varepsilon_0}{2} \Big[\mathbf{E}^2 + c^2 \mathbf{B}^2\Big] = 
	\displaystyle \frac{1}{2} \Big[ \frac{1}{\varepsilon_0}\mathbf{\Pi}^2 + \varepsilon_0c^2(\nabla\times \mathbf{A})^2\Big],
\end{aligned}
\end{multline}
where $c$ is the speed of light, and $\varepsilon_0$ is the dielectric permittivity of free space.

The axion part of the system for a laboratory size experiment may be represented by the uniform field $\theta$ and its canonical conjugate $\phi$:
\begin{multline}
\begin{aligned}
	\label{C004CO}
	\displaystyle  \mathcal{H}_\text{a} = \frac{\phi^2}{2m_a}+V(\theta),
\end{aligned}
\end{multline}
where $m_a$ is axion mass. Choosing the normal harmonic potential $V(\theta) = \frac{m_a\omega_a^2}{2}\theta^2$, this Hamiltonian reduces to a simple harmonic oscillator. On the other hand, for an experiment of reasonable size and duration, this representation is excessive; axion dynamics cannot be observed as the associated time constant should be extremely large compared to the experimental duration. The apparent finite `quality factor' of axion-induced photon signal is due to the velocity distribution of axion dark matter in the galactic halo, rather than due to response time of an `axion mode'. In this case, as it is usually done, the axion part is simply represented as an external signal $\theta$ with amplitude $\Theta$ and angular frequency $\omega_a$ that can be varied in time due to axion velocity.

It is widely accepted that the axion-photon interaction part in a laboratory size experiment can be represented in the following form:
\begin{multline}
\begin{aligned}
	\label{C003CO}
	\displaystyle  \mathcal{H}_{\text{int}} = \varepsilon_0 c g_{a\gamma\gamma} \theta \ \mathbf{E}\cdot \mathbf{B}  = -\frac{\varepsilon_0 c g_{a\gamma\gamma} \theta}{\varepsilon_0} \mathbf{\Pi}\cdot (\nabla\times \mathbf{A}),
\end{aligned}
\end{multline}
where $\theta$ is a scalar axion-like field, and $g_{a\gamma\gamma}$ is the commonly presented axion-photon coupling constant. 



\section{Two Modes Axion Electrodynamics}

We consider an electromagnetic cavity with two modes of two angular frequencies $\omega_1$ and $\omega_2$. Given a resonant structure, each mode $n$ is characterised by a certain distribution of electric ($\mathbf{E}_n(\mathbf{r})$) and magnetic field ($\mathbf{B}_n(\mathbf{r})$) in a certain finite volume:
\begin{multline}
\begin{aligned}
	\label{UI901R}
	\displaystyle  \mathbf{E}_n(\mathbf{r}) = -\frac{1}{\varepsilon_0}\Pi_n \mathbf{u}_n(\mathbf{r}) = iE_{V,n}(c_n-c_n^\dagger)\mathbf{e}_n(\mathbf{r}),\\
	\displaystyle  \mathbf{B}_n(\mathbf{r}) =  A_i \nabla\times\mathbf{u}_n(\mathbf{r}) = \frac{1}{c}E_{V,n}(c_n+c_n^\dagger)\mathbf{b}_n(\mathbf{r}),\\
\end{aligned}
\end{multline}
where $c_n^\dagger$ ($c_n$) are creation (annihilation) operators for the mode $n$, $E_{V,n} = \sqrt{\frac{\hbar\omega_n}{2\varepsilon_0 V_n}}$, and $\mathbf{e}_n(\mathbf{r})$ and $\mathbf{b}_n(\mathbf{r})$ are unit vectors representing the mode polarization.

Each of these two photonic modes is coupled to an axion signal via the term in Eq.~(\ref{C003CO}) that can be rewritten as (moving from Hamiltonian density to Hamiltonian):
\begin{multline}
\begin{aligned}
	\label{interaction}
	\displaystyle  H_\text{int} = \varepsilon_0 c g_{a\gamma\gamma} \theta \Big(\int_Vd^3r\mathbf{E}_1\cdot \mathbf{B}_2 + \int_Vd^3r\mathbf{E}_2\cdot \mathbf{B}_1\Big),\\
\end{aligned}
\end{multline}
where it is assumed that material properties dictate that for each mode $\mathbf{E}_i\cdot \mathbf{B}_i = 0$.
Quantising the two modes, one arrives at the following interaction Hamiltonian in terms of creation-annihilation operators:
\begin{multline}
\begin{aligned}
	\label{C007CO}
	\displaystyle  H_\text{int}  = i\frac{\hbar g_{a\gamma\gamma} \theta}{2} \sqrt{{\omega_1\omega_2}}\Big[\xi_1(c_2+c_2^\dagger)(c_1^\dagger-c_1) +\\
	 \displaystyle \xi_2(c_1+c_1^\dagger)(c_2^\dagger-c_2)\Big]\\
\end{aligned}
\end{multline}
where two dimensionless coefficients $\xi_1$ and $\xi_2$ represent overlap between the two modes:
\begin{multline}
\begin{aligned}
	\label{C008CO}
	\displaystyle \xi_1 = \frac{1}{\sqrt{V_1V_2}} \int_Vd^3r(\mathbf{e}_1\cdot \mathbf{b}_2),\\
	\displaystyle \xi_2 = \frac{1}{\sqrt{V_1V_2}} \int_Vd^3r(\mathbf{e}_2\cdot \mathbf{b}_1).
\end{aligned}
\end{multline}
These coefficients can span from one, when two modes are of the same shape and fully orthogonal, to zero when they exhibit no overlap. 
Finally, it is possible to separate swapping and parametric parts:
\begin{multline}
\begin{aligned}
	\label{C009CO}
	\displaystyle  H_\text{int}  = i\hbar g_\text{eff} \theta \Big[\xi_-(c_1c_2^\dagger-c_1^\dagger c_2) + \xi_+(c_1^\dagger c_2^\dagger-c_1 c_2)\Big],\\
\end{aligned}
\end{multline}
where $\xi_\pm=\xi_1\pm\xi_2$, and $g_\text{eff} = \frac{ g_{a\gamma\gamma}}{2}\sqrt{{\omega_1\omega_2}}$ is the effective trilinear coupling. 

A particular type of dynamics described by the interaction Hamiltonian (\ref{C007CO}) depends on the resonant frequencies of the photon modes $\omega_i$ and axion signal $\omega_a$. In particular, the following regimes can be identified (assuming $\omega_2-\omega_1 > 0$):
\begin{multline}
\begin{aligned}
	\label{C009CO}
	\displaystyle  \omega_a = \omega_2 + \omega_1, \text{axion downconversion}\\
	\displaystyle  \omega_a = \omega_2 - \omega_1, \text{axion upconversion}\\
\end{aligned}
\end{multline}
where neither of the angular frequencies is zero. A case where one of cavity frequencies is zero would correspond to the Sikivie detector\cite{Sikivie:1983aa} with virtual photons representing one of the modes. This case has been widely studied theoretically with a few experimental realisations in different frequency ranges\cite{Asztalos:2010aa,ORGAN}.

\begin{figure}
\includegraphics[width=0.65\columnwidth]{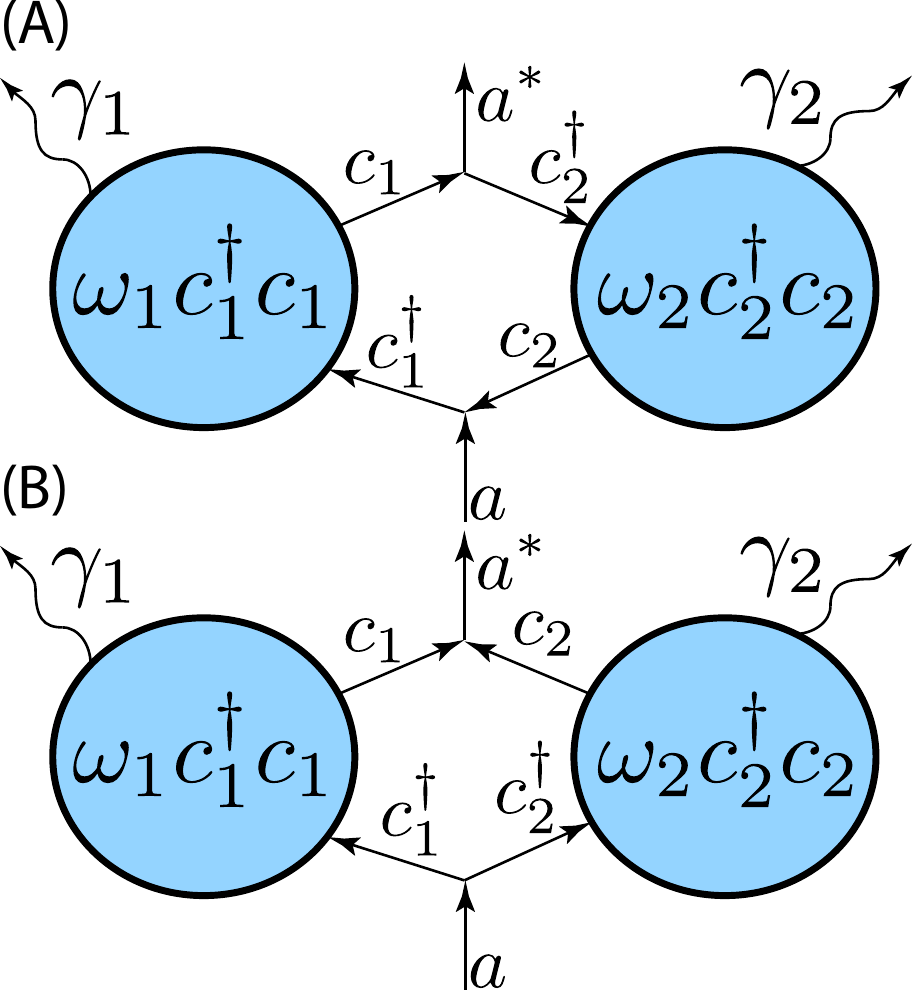}
\caption{Graphical representation of the two modes interacting through axion coupling in (A) upconversion and (B) downconversion cases.}
\label{diagram1}
\end{figure}

To demonstrate the difference between the upconversion and down conversion cases, we transform the system into the rotating frame associated with the frequencies of both modes and apply the rotating wave approximation (RWA). For this purpose, the axion signal $\theta$ may be decomposed in terms of complex amplitudes $a^\ast \exp(i\omega_a t)$ and $a \exp(-i\omega_a t)$. Conditions (\ref{C009CO}) give the two following Hamiltonians after corresponding RWA when fast rotating terms are removed (all couplings and signals are small as dictated by the weak signal detection problem):
\begin{multline}
\begin{aligned}
	\label{C010CO}
	\displaystyle  H_\text{D} = {i\hbar g_\text{eff}\xi_+}(a c_1^\dagger c_2^\dagger - a^\ast c_1 c_2),\\
	\displaystyle  H_\text{U} = {i\hbar g_\text{eff}\xi_-}(a^\ast c_1 c_2^\dagger - a c_1^\dagger c_2),\\
\end{aligned}
\end{multline}
in the corresponding interaction pictures. Graphical representations of these processes is given in Fig.~\ref{diagram1}. In both cases, the problem is reduced to a system of two piecewise coupled modes, as $a$ may be understood as a complex coefficient. The downconversion case Hamiltonian represents the parametric interaction found, for example, in the blue side band regime of optomechanical systems\cite{RevModPhys.86.1391}.The upconversion case is a swapping (or beam splitter) interaction that corresponds to the red sideband regime in optomechanics\cite{RevModPhys.86.1391}. For the upconversion and downconversion cases, the corresponding Heisenberg equations of motion are respectively:
\begin{multline}
\begin{aligned}
	\label{C410CO}
	\displaystyle  \frac{d}{dt}c_i = -{i}\omega_i c_i -\gamma_i c_i - g_\text{eff}\xi_- a c_j - {i}\sqrt{2\gamma_i}b^{\text{in}}_i,\\
	\displaystyle  \frac{d}{dt}c_i = -{i}\omega_i c_i - \gamma_i c_i + g_\text{eff}\xi_+ a c_j^\dagger - {i}\sqrt{2\gamma_i}b^{\text{in}}_i,\\
\end{aligned}
\end{multline}
where $j\neq i$, $\gamma_i$ and $b^{\text{in}}_i$ are a loss rate and input signal for the $n$th mode respectively.

Given the systems described by the Hamiltonians (\ref{C010CO}), one can imagine two classes of detection strategies. The first class relies on excess power detection introduced by the axion-photon coupling. This approach, further discussed in Section~\ref{PD}, is an extension to standard axion detectors using strong DC magnetic fields\cite{Sikivie:1983aa,Asztalos:2010aa,ORGAN} where the role of the DC field is played by the modes' oscillating fields. Besides the numerical values for the physically achievable field strengths, the major difference between the current proposal and the standard virtual field technique is the appearance of phases, making the experiment phase sensitive. It is worth noting that the Hamiltonian for the downconversion case may be interpreted as a trilinear Hamiltonian coupling pumped cavities to an external signal. This Hamiltonian has been considered previously in the literature in different physical contexts\cite{Nation:2010aa} including its application to parametric amplification\cite{Agrawal:1974aa, Gambini:1977aa}. This property might be exploited to increase the signal-to-noise ratio of the corresponding axion detector. 
The second class of strategies is related to measuring photon mode frequency shifts introduced by the axion mediated coupling terms (\ref{C010CO}) and has previously been discussed in the context of other hidden sector particle searches{\cite{Parker:2013aa}. In this approach, discussed in detail in Section~\ref{FS}, instead of detecting power coming from the modes, one is looking for mode frequency deviations associated with the new physics.

\section{Power Detection Approach}
\label{PD}

The excess power detection method is the most widespread approach to weak signal detection including the axion and paraphoton searches. According to this paradigm, one is looking for extra power coming from an experimental setup over the expected noise floor. Typically the noise floor is associated with thermal fluctuations of a measured device, i.e. cavity, and noise temperature of the amplifying electronics. Thus, such a scheme is usually realised in a cryogenic environment minimising all thermal fluctuations. 

The same approach can be realised using the analysed double mode system. Although, the potential for external signal pumping of oscillating fields brings new possibilities to the problem. Indeed, the axion coupling may be regarded as a mixing term in the presence of strong external pumping resulting in certain analogies that can be drawn with parametric amplifiers\cite{Yurke:2006aa,Eichler:2014aa}, which have found numerous applications on the forefront of physics. Parametric amplifiers' phase sensitivity and related phenomena allow them to beat the quantum or thermal limit or any other physical constraints to achieve extraordinary levels of sensitivity. Such devices could be built on different physical principles, for example, nonlinearities from superconducting junctions, nonlinear crystals, mechanical resonators or magnon systems\cite{Castellanos-Beltran:2007aa,Kochetov:2016aa,Cerullo:2003aa, Olkhovets:2001aa,Bracher:2017aa}. To analyse such systems, it is customary to split the problem into two steps by splitting variables into nonlinear large pumped amplitudes and small fluctuations giving linearised equations of motions\cite{Yurke:2006aa}. In Appendix~\ref{uno}, we follow this line of thought to analyse the sensitivity of the externally pumped double mode axion detector in the downconversion and upconversion cases. The result (Eq.~(\ref{C017CO}) for the upconversion and Eq.~(\ref{C017FT}) for the downconversion) suggests that the power in one mode due to an axion signal is proportional to the power stored in the other mode. This result is in accordance with the standard DC magnetic field detector where the axion power at the cavity output is proportional to $B_{\text{DC}}^2$, squared magnetic field strength, i.e. magnetic field energy. Regarding the amplitudes, rather than powers, the main difference is additional phase $-\phi_m$ (upconversion) or $+\phi_m$ (downconversion) that is not measurable with a standard power detector\cite{RfSikivie}. 
It is worth noting that the resulting spectral density from Eq.~(\ref{C017CO}) is identical to the previously proposed method in the case\cite{RfSikivie} when one mode is pumped on resonance and the signal is observed through the other mode. In practice, direct power detection using the double mode approach would be inferior to the scheme involving virtual photons (DC magnetic field) due to practical impossibility to create oscillating fields of matching strength.   

Instead of measuring the power of fluctuations from a single mode, a better strategy is to measure cross-correlation between the two modes of the system. Assuming statistical independence of the field fluctuations coming into the modes $\widetilde{b}_n^{\text{in}}$, the cross correlation spectrum of both channels is reduced to:
\begin{multline}
\begin{aligned}
	\label{C018CO}
	\displaystyle  S_{12}^{\text{D/U}}[\Omega] = \frac{g_\text{eff}^2\xi_\pm^2C_1C_2 }{|(i\Omega-\Gamma_1)(i\Omega-\Gamma_2)|}e^{i(\pm\phi_2\mp\phi_1)} S_{a}[\Omega],
\end{aligned}
\end{multline}
where upconversion and downconversion cases are different only through the phase factors giving the identical cross-correlation spectrum (\ref{C018CO}) with a different multiplier $\xi_+\rightarrow\xi_-$ due to different mode overlapping integrals (\ref{C008CO}). It means that by observing the power, the two cases are indistinguishable. One consequence of this is that by a single measurement, one searches for axions in both the upconversion and downconversion regimes. The sensitivities of these cases can vary because of the different overlap integrals. In case of a candidate detection, one would need either to modify geometry and thus $\xi_\pm$ or verify the two cases separately with a different combination of frequencies $\omega_1$ and $\omega_2$.



\section{Frequency Measurement Approach}
\label{FS}

Frequency measurement techniques have found considerable attention in precision sensing technology.  They have been successfully applied in such fields such as particle detection, bio-sensing, magnetic field and mass sensors, etc\cite{Frank:2012aa,Du:2017aa,Foreman:2015aa}, as well fundamental physics tests\cite{Parker:2013aa,Nagel:2015aa,Lo:2016aa,Goryachev:2018aa}. Unlike the power detection scheme, the fundamental limit in frequency measurements comes from the best achievable frequency stability of the measurement parts and thus it is not directly related to the Nyquist noise and the ambient temperature, although cryogenic cooling may significantly improve the detection limit due to higher quality factors. It is generally accepted that one can detect frequency variations on the order of $10^{-6}$ relative to the system linewidth. With such systems as superconducting cavities and sapphire, it is possible to achieve linewidths of the order of 1 Hz at microwave frequencies\cite{Locke:2003aa} giving unprecedented sensitivity of the frequency measurement approach. 

In this section we consider a possible application of frequency metrology to axion searches enabled by the proposed dual mode approach. As before, we split the discussion into the downconversion and upconversion parts.

\subsection{Axion Induced DC Frequency Shifts}

The system equations of motion in the interaction picture for the downconversion (\ref{C009CO}) can be written in the following form:
\begin{multline}
\begin{aligned}
	\label{C011XY}
	\displaystyle  \frac{d}{dt}C_1 =  (-\gamma_1-i\Delta_1)c_1+g_\text{eff}\xi_+ A C_2^\dagger,\\
	\displaystyle  \frac{d}{dt}C_2 = (-\gamma_2-i\Delta_2) c_2+g_\text{eff}\xi_+ A C_1^\dagger,\\
\end{aligned}
\end{multline}
where $C_1$ and $C_2$ are slowly varying amplitudes, $\gamma_n$ is the loss rate for the mode $n$, $\Delta_n$ is the detuning frequency of a cavity with angular frequency $\overline{\omega}_i$ such that $\Delta_1 + \Delta_2 = \overline{\omega}_1 + \overline{\omega}_2 - \omega_a$, and $A$ is the axion field. Introducing $\Gamma_n = \Delta_n-i\gamma_n$, the eigenvalues of this system are
\begin{multline}
\begin{aligned}
	\label{C012XY}
	\displaystyle  e_\pm^\text{D} = \frac{\Gamma_1+\Gamma_2^\ast}{2} \pm\frac{1}{2}\sqrt{(\Gamma_1-\Gamma_2^\ast)^2+4g_\text{eff}^2\xi_+^2|A|^{2}} \\
	\displaystyle = \Delta_+-{i\gamma_-}\pm\sqrt{(\Delta_--{i\gamma_+})^2+g_\text{eff}^2\xi_+^2|A|^{2}}
\end{aligned}
\end{multline}
as well as their complex conjugates, where $\gamma_\pm = (\gamma_1\pm\gamma_2)/2$ and $\Delta_\pm = (\Delta_1\pm\Delta_2)/2$.

For the upconversion case, the equations of motion could be written as follows:
\begin{multline}
\begin{aligned}
	\label{C013XY}
	\displaystyle  \frac{d}{dt}C_1 = (-\gamma_1-i\Delta_1) c_1-g_\text{eff}\xi_- A C_2,\\
	\displaystyle  \frac{d}{dt}C_2 = (-\gamma_2-i\Delta_2) c_2+g_\text{eff}\xi_- A^\ast C_1,\\
\end{aligned}
\end{multline}
with the eigenvalues:
\begin{multline}
\begin{aligned}
	\label{C014XY}
	\displaystyle  e_\pm^\text{U} = \frac{\Gamma_1+\Gamma_2}{2} \pm\frac{1}{2}\sqrt{(\Gamma_1-\Gamma_2)^2+4g_\text{eff}^2\xi_-^2|A|^{2}}\\
	\displaystyle = \Delta_+ -{i\gamma_+} \pm \sqrt{(\Delta_--{i\gamma_-})^2+g_\text{eff}^2\xi_-^2|A|^{2}}
\end{aligned}
\end{multline}
and their complex conjugate.

In the simplest and most sensitive form, one tunes both detuning frequencies to zero ($\Delta_\pm=0$) and match the losses ($\gamma_1=\gamma_2$). The resulting relative shift of eigenfrequencies due to coupling to axion signal of amplitude $|A|$ is
\begin{multline}
\begin{aligned}
	\label{C015XY}
	\displaystyle  \partial e^\text{D} =e^\text{D}_+-e^\text{D}_-= \sqrt{g_{a\gamma\gamma}^2\omega_1\omega_2\xi_+^2|A|^{2}-\gamma^2_+},\\
	\displaystyle  \partial e^\text{U} =e^\text{U}_+-e^\text{U}_-= g_{a\gamma\gamma}\sqrt{\omega_1\omega_2}\xi_-|A|.
\end{aligned}
\end{multline}
Whereas for the upconversion case, splitting is directly proportional to the axion amplitude, in the downconversion case only imaginary part is sensitive to the presence of axions. 
To compare the two cases of detection in more detail, we calculate deviation of real and imaginary components of the eigenvalues from its values in the axion-free case ($|A| = 0$) as a function of the normalised axion coupling strength $\chi = \frac{g_\text{eff}\xi_\pm|A|}{\gamma_+}$ shown in Fig.~\ref{sensitivity1}. The upconversion case (dashed line) demonstrates the expected linear dependence of the real component of eigenfrequency deviation on the coupling for small $\gamma_-$ and $\Delta_-$ as stated by Eq.~(\ref{C015XY}). This regime demonstrates eigenfrequencies shifts equal to axion signal strength normalised to frequency units. This case requires precise matching of the modes in terms of losses and detuning frequencies. With nonzero difference between mode losses $\gamma_-$ the sensitivity in the lower coupling limit decreases and is limited only to the imaginary part of the eigenvalue. 

\begin{figure}
\includegraphics[width=1\columnwidth]{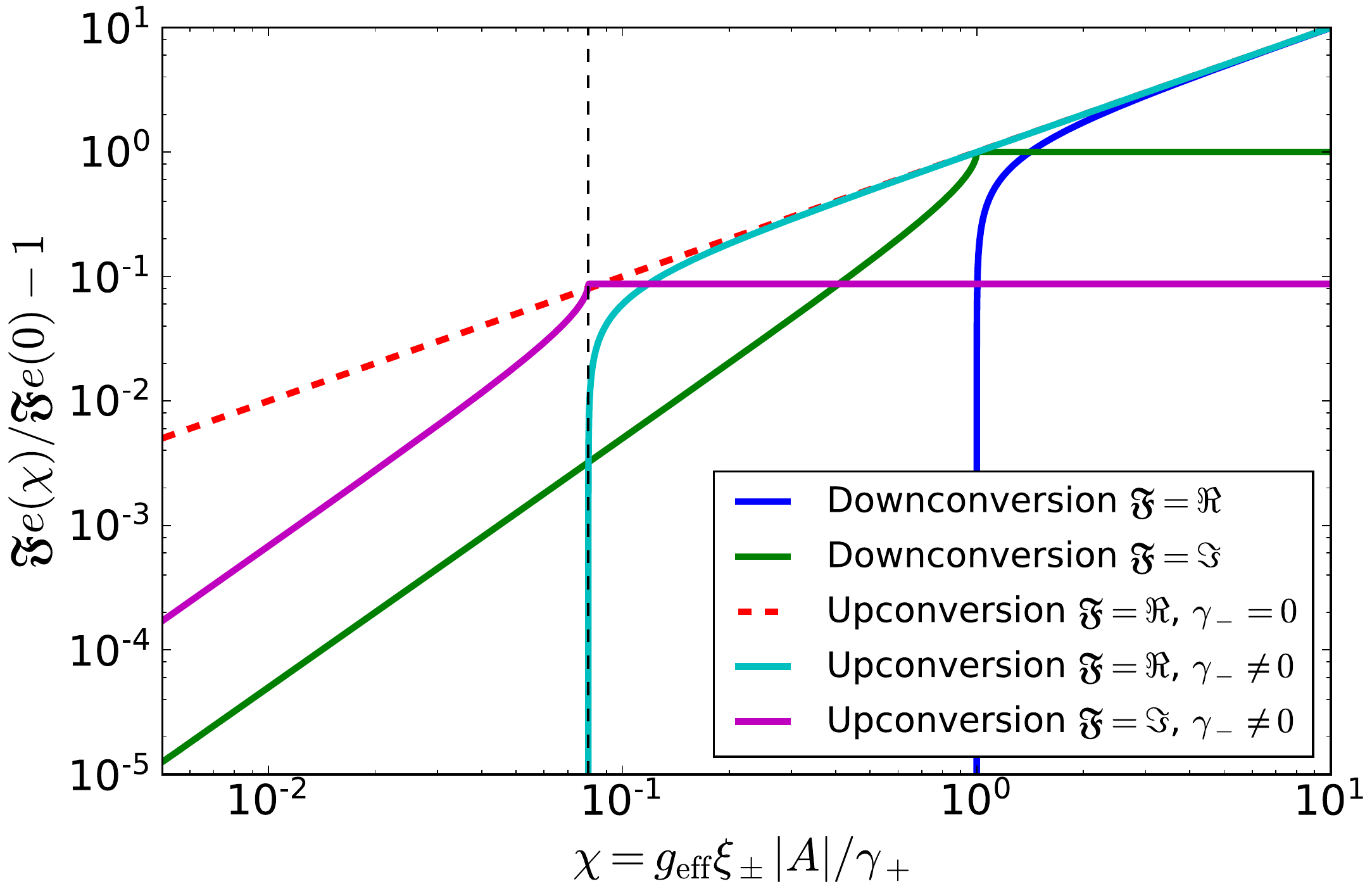}
\caption{DC sensitivity of the dual mode detector eigenvalues to the normalised axion coupling strength $\chi$.}
\label{sensitivity1}
\end{figure}

In the downconversion case, although the sensitivity is reduced, the axion coupling appears as a modification to the imaginary part of eigenfrequencies: the linewidth gets narrower when coupling increases. The change in the imaginary part is shown with the green curve in Fig.~\ref{sensitivity1}. At $\chi = 1$, the axion term balances the sum of the losses leading to the steady state oscillation regime. In this regime, photons created by axion downconversion, balance the cavity losses $\gamma_+$ and the system starts oscillating. Since the axion signal is extremely weak, this regime is not achievable even for the most narrow-linewidth cavities. In order to boost the sensitivity of the frequency measurement technique in the downconversion case, one may look for compensation of the cavity losses.Indeed, by introducing external or internal gain into one or two modes, the effective cumulative losses $\gamma_+$ decrease resulting in lower oscillation threshold. 

\subsection{Open Loop Axion Induced Spectral Density of Phase Measurements}

In practice, measurement of DC frequency shifts is technically challenging. For fundamental tests the situation is worsened by the requirement to verify or veto the candidate signals. The problems may be solved by modulating one of the system parameters as, for example, proposed in the case of a paraphoton search\cite{Parker:2013aa}. Another approach is to search for axion signals in the Fourier spectrum. For such an approach, we consider a regime of small axion detuning: $\omega_a = \omega_1\pm\omega_2+2\pi f$ where $2\pi f \ll \omega_1$. Note that these relations do not violate the energy conservation, but rather indicate the regime in which the electromagnetic modes are slightly detuned from the axion frequency. Such a regime is analogous to the situation in which the cavity mode frequency of the standard DC haloscope is slightly shifted from the axion frequency. The detuning frequency $f$ plays the role of the Fourier spectrum in the generated noise and could span over a few decades. In this situation, the axion amplitude appears as a slowly varying parameter $A$ in the EOMs (\ref{C011XY}) and (\ref{C013XY}). The detailed analysis of this approach is given in Appendix~\ref{A1} where we reformulate the problem in terms of slowly varying real magnitude and phase\cite{Rubiola:2008aa,Goryachev:2011aa,Gbook} instead of the complex amplitude representation (\ref{C011XY}) and (\ref{C013XY}).

The analysis in Appendix~\ref{A1}, reveals a transfer function from one of the axion quadratures to the phase fluctuations of the output signal (Eq.(\ref{R007VB}) for upconversion and Eq.~(\ref{R104VB}) for downconversion). Using this transfer function, the phase noise spectrum of the output signal for the $i$th mode may be represented as follows:
\begin{multline}
\begin{aligned}
	\label{TT001}
	\displaystyle  S_{\varphi,i}^{\text{D/U}}(f) =  \frac{g_\text{eff}^2\xi_\pm^2}{f^2+\gamma_i^2}\Big|\frac{\overline{x}_j}{\overline{x}_i}\Big|^2 S_A(f) + \frac{\gamma_i^2}{f^2+\gamma_i^2}S_\theta(f),
\end{aligned}
\end{multline}
where the first component is the spectrum of phase induced by the axion signal, and $S_A(f)$ is the power spectrum of the axion field, in units of $\left|A\right|^2$, where $A$ represents the axion field, and the second term is due to technical phase fluctuations $S_\theta(f)$ of the pump signal. The result explicitly depends on the ratio of steady state amplitudes $\overline{x}_i$ in both modes. Thus, the overall sensitivity may be boosted using this ratio. It is also important to note that despite the fat that the axion signal is filtered by the resonator (appears as the first order transfer function in the phase space), the overall signal-to-noise ratio is constant as the technical fluctuations are also filtered by the same filtering function. 

The phase noise spectrum (\ref{TT001}) could be measured using the phase measurement setup shown in Fig.~\ref{scheme1}.  Here, the output of each cavity is mixed with a local oscillator of the same frequency as the pump. By varying the phase between the pump signal incident on the cavity and the mixer $\theta$, one can access to both quadratures of the field fluctuations in the cavity. Such an approach is capable in principle of detecting phase fluctuations in rms amplitude of $2\times10^{-11}$rad$/\sqrt{\text{Hz}}$ at Fourier frequencies above a few kiloHertz\cite{Ivanov:2009aa}. Such kind of measurement is not possible with the traditional DC-field axion detector. 

\begin{figure}
\includegraphics[width=1\columnwidth]{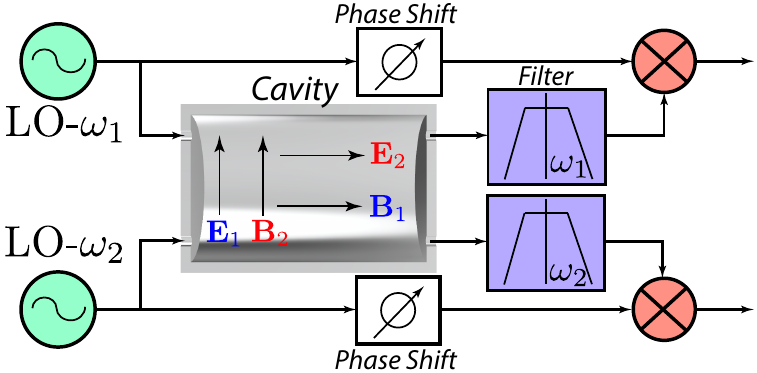}
\caption{An open loop realisation of the dual mode axion detection scheme. Two external local oscillators (LO) are used to excite two axion coupled modes.}
\label{scheme1}
\end{figure}

The sensitivity plots of the pumped phase noise measurement experiment is shown in Fig.~\ref{openloop}. These sensitivities are based on comparing the size of the axion-induced phase-shifts, which are taken to be of magnitude 
\begin{equation*}
\sqrt{S_{\varphi,i,a}^\text{D/U}}=\sqrt{\frac{1}{f^2+\gamma_i^2}}\frac{g_{a\gamma\gamma}\sqrt{\omega_1\omega_2}}{2}\xi_\pm\Big|\frac{\overline{x}_j}{\overline{x}_i}\Big|\left|A\right|,
\end{equation*}
with the minimum detectable phase-shift $\frac{\sqrt{S_\phi(f)}}{\sqrt{t}}$ where in this case the background noise, $S_\phi(f)$ is the phase noise spectral $\emph{density}$ of the output we are measuring (as opposed to the phase noise $\emph{spectrum}$, $S_\theta$), and $t$ is the averaging time~\cite{LIPhaseNoise}.

The green and blue lines in fig.~\ref{openloop} represent the potential sensitivity to the up and downconversion regimes of this kind of experiment. The pump signal is taken to originate from a state of the art, frequency stabilized cryogenic sapphire oscillator (CSO) operating at the noise floor of a partially cryogenic frequency discrimination system~\cite{LONoiseFloor} and white noise background. We further assume that the ratio of powers in the two modes is 1000, such that the readout mode has 1000 times less power dissipated in it. The mode structures, quality factors, resonance details and overlap integrals are discussed in section~\ref{Design}. Generally speaking, such a search would operate by detuning the two modes by some distance in frequency space, and then searching the Fourier phase-noise spectrum of the one of the resonances (say the higher frequency resonance for instance) for peaks corresponding to axion-induced phase-shifts.

Of course, for a given resonator geometry the upconversion technique will never be able to reach the same high-mass range as the downconversion technique, and so the extension of the green line into the same frequency range as the blue line in fig.~\ref{openloop} represents the sensitivity if we were to construct some resonator with the same sensitivity, but with mode frequencies such that this range was achievable with the upconversion technique. This is presented for the purpose of the direct comparison of the two techniques.
By searching a few MHz in Fourier space for such peaks, we are sensitive to slightly detuned axions such that $\omega_a = \omega_1\pm\omega_2+2\pi f$, where $f$ is our range of Fourier frequencies. We may then further detune the resonances and repeat the process, gradually excluding a large section of the axion mass range. Of course, averaging for longer can yield improvements, and the above plots are based on an averaging time of 30 days per 100 MHz. The amount of frequency space covered in that 30 day period will then depend on the range of Fourier frequencies measured.

\begin{figure}
\includegraphics[width=1\columnwidth]{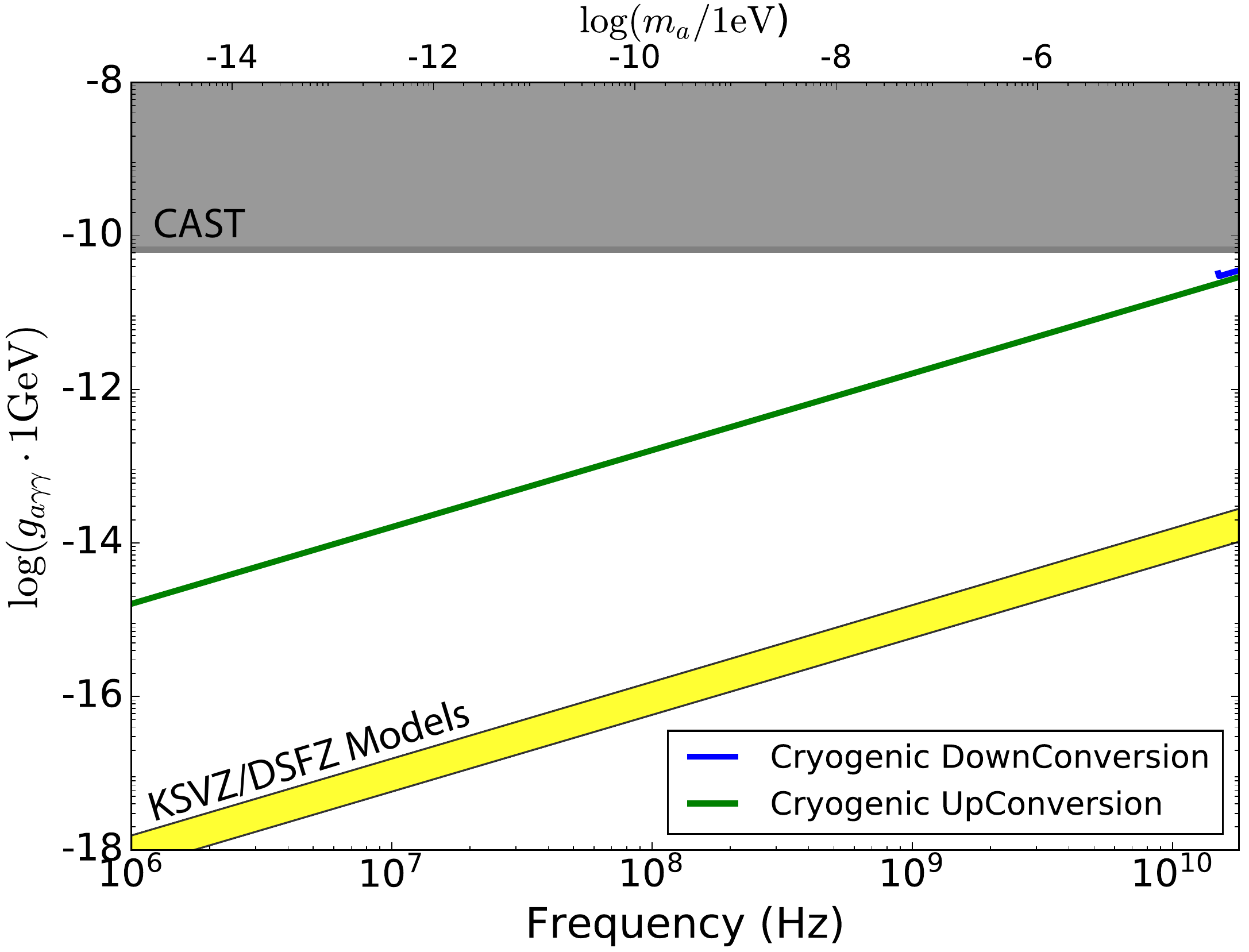}
\caption{Sensitivity of the cavity phase noise measurement experiment comparing to the axion models and existing limits due to the CAST experiment. }
\label{openloop}
\end{figure} 

\subsection{Loop Oscillator Axion Induced  Spectral Density of Phase Measurements}

Instead of relaying on an external frequency source, one may construct an oscillator using the double mode cavity as a frequency selective element. Such a measurement setup is shown in Fig.~\ref{scheme2}. Discussion of such a system is given in Appendix~\ref{oscil} where the spectrum of phase fluctuations is given as follows:
\begin{multline}
\begin{aligned}
	\label{TT0011}
	\displaystyle  S_{\varphi,i}^{\text{D/U}}(f) =  \Big[1+\frac{\gamma_i^2}{f^2}\Big]\Big(\frac{g_\text{eff}^2\xi_\pm^2}{f^2+\gamma_i^2}\Big|\frac{\overline{x}_j}{\overline{x}_i}\Big|^2 S_A(f) +S_\theta(f)\Big),
\end{aligned}
\end{multline}
where $S_\theta$ are technical fluctuations inside the oscillator loop, and $S_A(f)$ is defined as in (\ref{TT001}). The multiplying factor in this result is due to the Leeson effect\cite{leeson}.

\begin{figure}
\includegraphics[width=1\columnwidth]{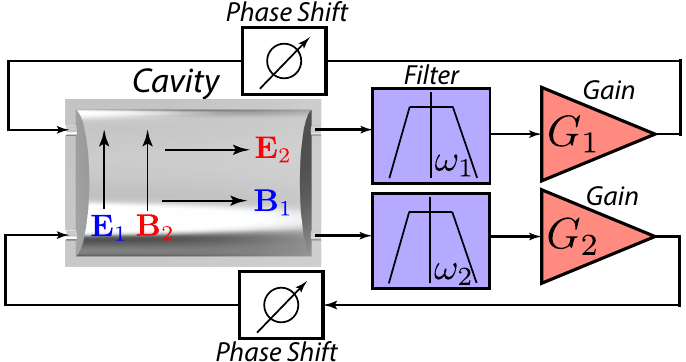}
\caption{Loop oscillator approach to dual mode axion detection. Two positive feedback loops utilise axion coupled modes as frequency selective elements.}
\label{scheme2}
\end{figure}

The sensitivity plots of the loop oscillator experiment for the room temperature and cryogenic implementations are shown in Fig.~\ref{oscillators}. These sensitivities are calculated in a similar fashion to the externally pumped loop experiment above. Indeed, the experiment would operate in the same way, by detuning the resonances some amount and searching Fourier space, before detuning the resonance further and repeating the search. We again compare the magnitude of axion-induced phase shifts - in this case given by 
\begin{equation*}
\sqrt{S_{\varphi,i,a}^\text{D/U}}=\sqrt{\left(1+\frac{\gamma_i^2}{f^2}\right)\left(\frac{1}{f^2+\gamma_i^2}\right)}g_{a\gamma\gamma}\sqrt{\omega_1\omega_2}\xi_\pm\Big|\frac{\overline{x}_j}{\overline{x}_i}\Big|\left|A\right|,
\end{equation*}
with the minimum detectable phase shift $\frac{\sqrt{S_\phi(f)}}{\sqrt{t}}$ where in this case the background noise, $S_\phi(f)$ is the phase noise spectral density of the loop oscillator that we are measuring the output of (as opposed to the phase noise $\emph{spectrum}$, $S_\theta$), and $t$ is the averaging time.

For the cryogenic case, we are again assuming that the oscillator is a state of the art CSO operating at the noise floor of a partially cryogenic frequency discrimination system and white noise background, and for the room temperature measurements we assume frequency stabilized loop oscillators based on copper cavities with the same properties as the open loop experiment, again discussed in more detail in section~\ref{Design}, and amplifiers with effective added noise temperatures of 50 K, dissipating 1 W in the resonator, operating at the noise floor of the frequency discriminator and white noise background. It is important to note that for the upconversion experiment the sensitivities linearly decrease down to very low values of the frequency. 

\begin{figure}
\includegraphics[width=1\columnwidth]{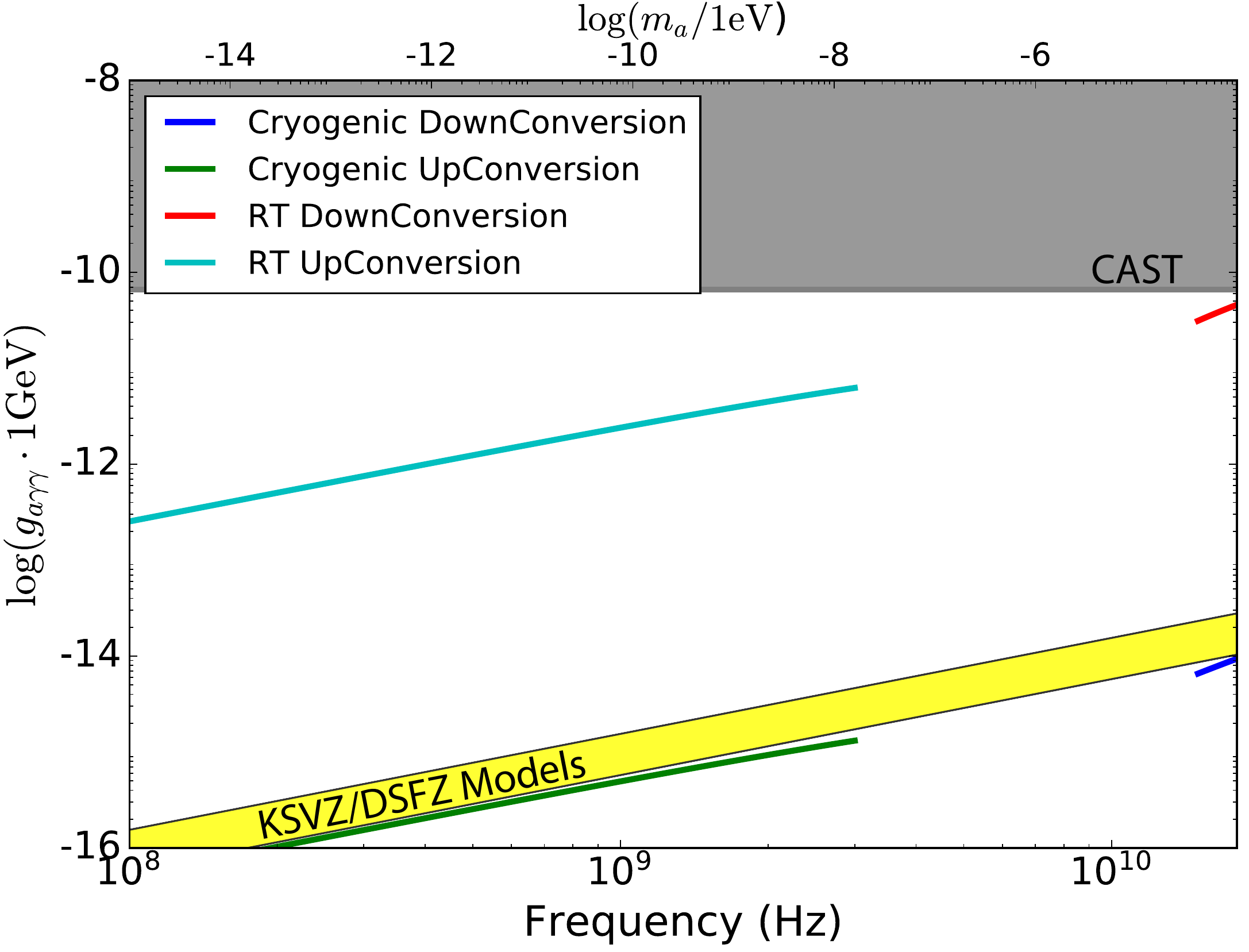}
\caption{Sensitivity of the room temperature and cryogenic versions of the loop oscillator experiment comparing to the axion models and existing limits due to the CAST experiment. The dashed line is the extension of the upconversion non-degenerate case down to low frequencies. }
\label{oscillators}
\end{figure} 

In addition to purely frequency and phase measurement schemes presented in Fig.~\ref{scheme1} and Fig.~\ref{scheme2}, one may consider different hybrid implementations. For example, one mode is used with a feedback loop as an oscillator, and the other is employed to measure the axion coupling induced phase shift with a help of external (highly stable) frequency source.

\section{Degenerate Mode Broadband Experiment}

An interesting class of detection techniques may be developed based on the degenerate frequency case scenario, where we require the two modes have equal frequencies ($\omega_1 = \omega_2 = \omega$), and axion frequency (mass) is small ($\omega_a\ll\omega$). Under these requirements, only the upconversion case is feasible. 
In the degenerate frequency case, considering only pure tones would put an exclusive requirement on the axion mass being near zero. The actual signal would be searched for in a wide range of offset (Fourier) frequencies as discussed in Section~\ref{FS} and developed in Appendices~\ref{A1} and \ref{oscil}. The obvious advantage of this approach is its broadband nature. Indeed, by measuring phase and amplitude noise in the degenerate case, one has direct access to a few decades of Fourier frequencies. Such broadband measurements could be directly realised with the modern reconfigurable digitizers. 

To analyse the Equations of Motion in the degenerate case, one may apply the same logic as in Appendix~\ref{uno}, \ref{A1}, \ref{oscil}.
Thus, the results obtained in Section~\ref{PD} and \ref{FS} are valid for the degenerate case as well.

The possibility to design a cavity with two orthogonal modes of the same frequencies have been demonstrated before\cite{Shigeru:1995aa,Harvey:2003aa}. For the microwave frequency range, one may argue that the mode degeneracy is not achievable due to unavoidable imperfections of a cavity. Such imperfections introduce coupling between the two modes that results in avoided level crossings. And it is due to this phenomenon the modes never coincide in the frequency space. Though, this is true, it is feasible to match the frequencies within the mode bandwidth: by minimising the all sorts of "imperfections", e.g. strongly coupled external probes, one may reduce mode-mode coupling to a value that is much smaller than the mode bandwidth. In this case, no avoided level crossing could be observed. 
An example of a microwave cavity that can be used for the degenerate two mode axion sensing is discussed further in Section~\ref{Design}.

Another practical problem immediately encountered with the degenerate frequency measurement scheme is inability to apply filters and separate the two signals. For example, in the frequency detection scheme with two oscillators, the oscillators will synchronize due to unavoidable coupling between the two resonances of the same frequency. A possible solution to this problem is to treat the signals as one and use an external clock as a reference as depicted in Fig.~\ref{scheme3}. 

\begin{figure}
\includegraphics[width=1\columnwidth]{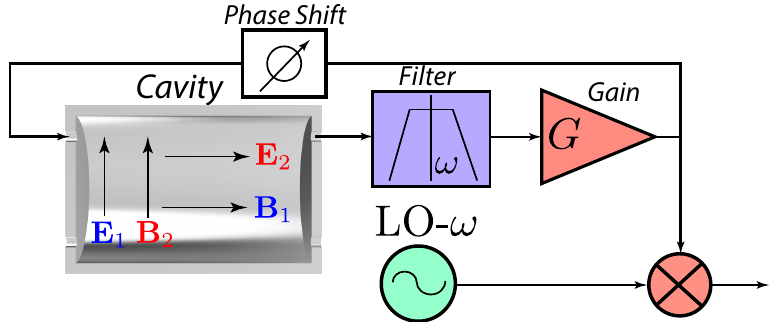}
\caption{A possible realisation of the degenerate mode detection scheme where two axion coupled modes of the same frequency are used as a frequency selective element for a feedback loop. The generated signal is compared against an external Local Oscillator (LO).}
\label{scheme3}
\end{figure}

Projected sensitivity of the degenerate broadband experiment operated at both room and cryogenic temperature ranges is shown in Fig.~\ref{broadband}. This plot is calculated based on the same parameters and phase expressions as the detuned loop oscillator experiment, simply setting the two resonance to the same frequency and sampling the phase noise spectrum up to 100 MHz. We again assume 30 days averaging time for this 100 MHz span  - this will be limited to some extent by the achievable frequency resolution of the digitiser. We note that these limits curve upwards as the loop oscillator phase noise hits the white phase noise floor some distance into the Fourier spectrum. As a result the straight, linear curves for the exclusion limits in fig.~\ref{oscillators} (and indeed, the dashed red line in fig.~\ref{broadband}) represent extending the flatter part of the ``broadband" exclusion limits, as it is in principle possible to achieve this level of sensitivity at any point in the accessible frequency space, by simply setting the detuning frequency accordingly and sampling only a narrow region of Fourier space. Such an experiment would naturally scan far slower than 30 days per 100 MHz, but the curves in figs.~\ref{broadband} and~\ref{oscillators} represent the limits of sensitivity achievable with room temperature and cryogenic loop oscillators, with feasible parameters, and within reasonable laboratory time scales. It is important to note that these schemes, and indeed the open loop scheme presented in fig.~\ref{openloop}, do carry technical difficulties, and will require careful design. For instance, these limits rely on the implementation of an improved frequency stabilisation system for a CSO. However, this is not beyond the realm of what is achievable, and the presented limits should beviewed as the potential reach of this technique.

\begin{figure}
\includegraphics[width=1\columnwidth]{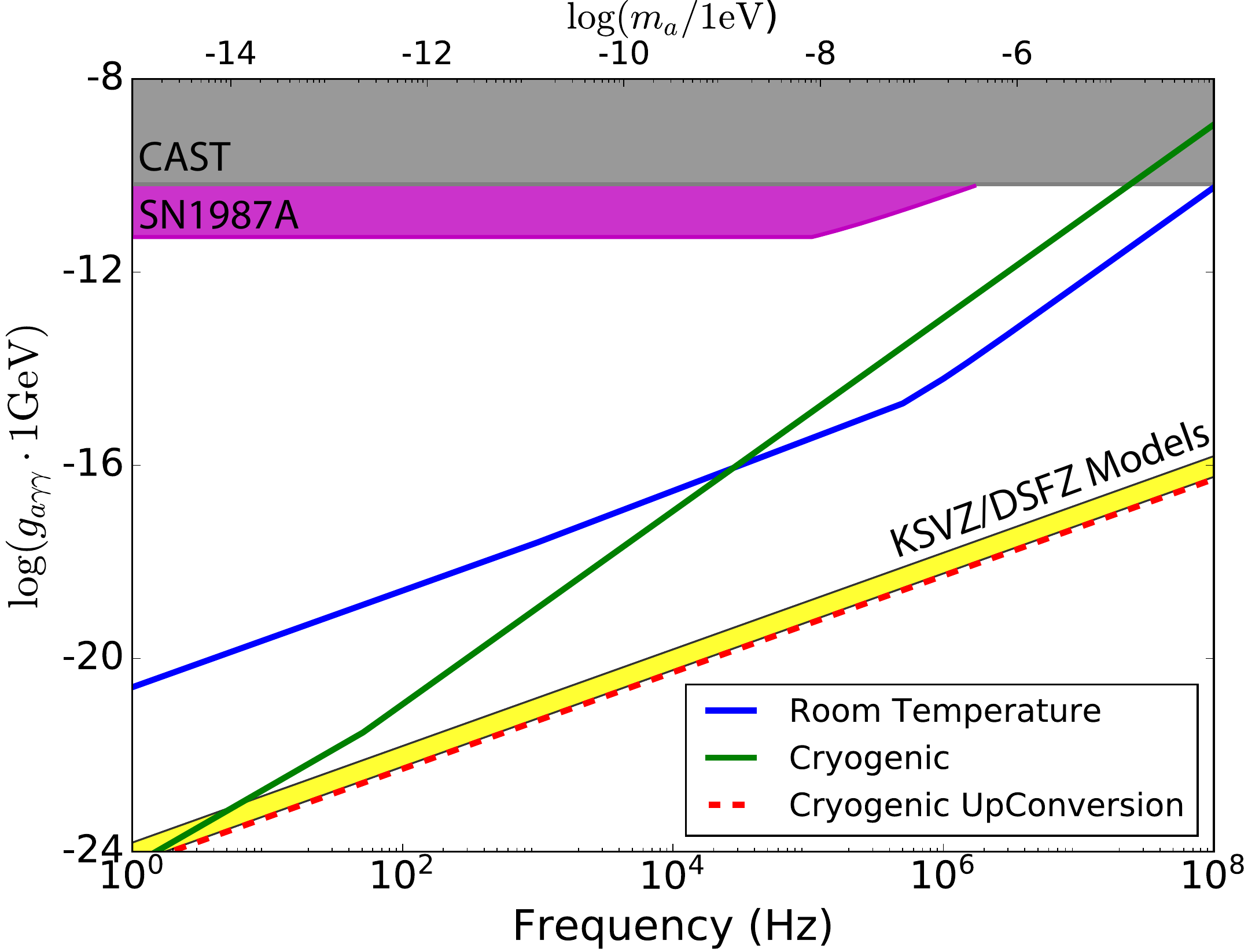}
\caption{Sensitivity of the room temperature and cryogenic versions of the degenerate mode experiment comparing to the axion models and existing limits due to CAST experiment and SN1987A.}
\label{broadband}
\end{figure}

\section{Some Practical Realisations}
\label{Design}

All dual mode techniques considered in the present work rely on a possibility to design a cavity that exhibits significant orthogonality of two modes represented by Eq.~(\ref{C008CO}). These formulae give the cavity form factor that has to be as large as possible. 

\subsection{TE-TM mode Cylindrical Cavity Resonator}

As discussed, the sensitivity limits above are calculated based on the example of a microwave cavity with two orthogonally polarized resonances. Many other geometries are possible, but for the purposes of demonstrating the sensitivity of these techniques we have modelled the following. We take a 29.2 mm radius cylindrical copper cavity with a TM$_{020}$ mode frequency of 9 GHz, and a TE$_{011}$ mode frequency tunable from 6.5 to 9 GHz as the cavity height tunes from 18.5 to 83.6 mm. The loaded quality factors of both resonances are taken to be 10,000, which is readily achievable in copper at these frequencies. In this configuration, taking the stationary frequency TM$_{020}$ mode to be mode ``2" in the equations for the overlap integrals, we find that $\xi_-$ varies from -0.39 to -0.50 as over the tuning range, whilst $\xi_+$ varies from 0.46 to 0.57. Generally speaking, the analytical expressions for these overlap integrals are messy and complex, and it is preferable to calculate them numerically for a given set of modes and cavity geometry. We found that the magnitude of these expressions generally increased with the cavity aspect ratio, $\frac{h}{a}$ where $h$ is the height of the cavity, and $a$ the radius, up to some maximum value achieved for aspect ratios greater than $\sim5$.

\subsection{Orthogonally Polarised Modes in a Fabry-P{\'e}rot Cavity}

Another interesting orthogonally polarized mode scheme to consider relies on Gaussian beam modes in Fabry-P{\'e}rot cavities. Gaussian beams have the advantage that the electric and magnetic fields have the same profile, simply rotated 90 degrees with respect to one another. If we take the electric field of a given Gaussian beam to be polarized on the $x$-direction and propagating in the z-direction such that
\begin{equation}
\vec{E}=E(r,z)~\mathbf{x},
\end{equation} 
then the magnetic field is
\begin{equation}
\vec{B}\propto E(r,z)~\mathbf{y}
\end{equation}This means, that if we were to take two orthogonally polarized Gaussian beam modes in the same Fabry-P{\'e}rot cavity, with the same frequency, the variables $\xi_+$ and $\xi_-$ would be 2 and 0 respectively. This has promising implications for high mass axion searches, via the downconversion technique.

However, we may also boost the sensitivity of the upconversion technique, by employing two modes orthogonally polarized in the same cavity, but of different frequency, ie a different number of wavelengths between the two mirrors. This would give a non-zero value of the $\xi_-$ parameter, which would open up the possibility of lower mass axion searches with such structures.

\section*{Conclusions}

We demonstrated that precision frequency and phase metrology could be used as a highly sensitive tool in dark matter detection. Because measurements are made in the space of frequencies and phases rather than in the space of amplitudes, the fundamental limit is set by control electronics rather than due to bare thermal fluctuations. This fact allows us to introduce a number of detection schemes matching or exceeding sensitivity of state-of-the-art cryogenic axion detectors. This includes a broad band low mass axion detection based on the degenerate case and Fourier spectra. In addition to bare axion amplitude detection, the method allows to deduce relative phase of the axion signal. The proposed new class of detectors may be understood as a phase sensitive detection scheme similar to phase sensitivity of parametric amplifiers. In summary, the main advantages of the proposed frequency control method are:
\begin{itemize}
\item magnet-free. Unlike traditional haloscopes\cite{ORGAN,ADMX2010,CAPP,CAPPToroid,MADMAX,HAYSTAC}, the proposed method does not require a strong DC magnetic fields;
\item SQUID-free. All sensitivities calculated in this work are based on usage of traditional low noise semiconductor amplifiers. Though superconducting technology might be used in the future, its presence is not crucial contrary to traditional metods\cite{ADMX2010};
\item cavity volume independence. Although cavity volume influences many parameters of the experiment such as resonance frequencies and quality factors, the sensitivity is not directly proportional to this parameter unlike in traditional haloscopes\cite{HAYSTAC,CAPPPIZZA,CAPPPhase,DielectricRing,AxionArray}. This removes the major obstacle for higher mass ($f_a>$10GHz) axion searches. Moreover, optical cavities might be used to probe otherwise unaccessible regions of THz and infrared spectrum as well as millimiter-wave and microwave frequencies;
\item Liquid-Helium temperature operation ($>4$K) where only a limited number of components such as cavity and amplifiers have to be at low temperature. This factor removes the need of dilution refrigeration that is a key component in traditional haloscopes\cite{ORGAN,ADMX2010,CAPP,CAPPToroid,HAYSTAC} making the whole experiment available to a broader audience. Although dilution refrigeration might give some incremental improvement in the axion search, all ultra-stable microwave and optical clocks and oscillators do not require temperatures below 4K.
\item access to higher and lower frequency ranges. The fact that actual axion mass is either the sum or difference of working frequencies opens a possibility to search for axions in less accessible frequency rages. For instance, working around 20GHz, one is able to probes axion masses in the vicinity of 40GHz, where experiments are significantly more difficult;
\item limited power levels ($P<100 \mu$W cryogenically and 1 W at room temperature). Although, the sensitivity does explicitly depend on power levels, on the current calculation only limited power levels are used unlike in some other proposals\cite{ALPS1,ALPS2,RfSikivie};
\item axion phase sensitive. Comparing to DC magnet haloscopes, the dual frequency method is able to provide additional information about axions, particularly its phase relative to pump signals, although that might lead to more complicated detection schemes;
\item KSVZ/DSFZ\cite{K79,SVZ80,DFS81,Z80} achievable. It is estimated that the cryogenic dual mode experiment is able to achieve the limit of the widely accepted axion dark matter models. On the other hand, even a tabletop search may lead to competitive limits on dark matter;
\item broadband search for low mass axions is possible\cite{ABRACADABRA,BEAST}. It is demonstrated that a wideband search that does not require tuning is possible.
\end{itemize}

\section*{Acknowledgements }

This work was supported by the Australian Research Council grant numbers DP160100253 and CE170100009.

\section*{References}

\appendix 

\section{Derivation of Two Mode Power Sensitivity}
\label{uno}

As it is mentioned in Section~\ref{PD}, the problem of axion detection with two pumped modes has much in common with the analysis of parametric amplifiers\cite{Yurke:2006aa,Eichler:2014aa}, where one is looking at manipulation of very weak signals on top of strong pumps.  Following the standard analysis in the field of parametric amplifiers, firstly, the steady state problem is solved for the strong pump signals. In the case of Josephson parametric amplifiers, these equations of motions are nonlinear. Secondly, the system dynamics is analysed for small fluctuations using linearised equations. For this reason, each field component is split into strong pumped parts ($C_n$ and $B_n^{\text{in}}$) and small fluctuations ($\widetilde{c}_n$ and $\widetilde{b}_n^{\text{in}}$):
\begin{multline}
\begin{aligned}
	\label{C013CO}
	\displaystyle  c_n = \big(C_n e^{-i\phi_n} + \widetilde{c}_n(t)\big)e^{-i\omega_{pn}t},\\
	\displaystyle  b_n^{\text{in}} = \big(B_n^{\text{in}} e^{-i\varphi_n} + \widetilde{b}^{\text{in}}_n(t)\big)e^{-i\omega_{pn}t},\\
	\displaystyle  a_n = \widetilde{a}(t)e^{-i\omega_{a}t},\\
\end{aligned}
\end{multline}
where $\omega_{pn}$ and $B^{\text{in}}_n$ represent pumping frequency and magnitude for the $n$th mode and the axion signal has only the small component $\widetilde{a}$. One may extend the model to include an additional output ``probe" and explicitly distinguish coupling and intrinsic losses. In the present analysis, $\widetilde{b}$ may be associated with technical fluctuations and noise, while $\widetilde{a}$ is the component that has to be detected. The latter quantity is assumed to be slowly time-varying to account for the "Quality factor" or linewidth of the axion signal that appears in many cosmological models.  Also, the phases of the pump signals $\varphi_n$ are introduced with respect to the axion phase that is set to zero.

On the first step, due to extreme weakness of axions, the axion mediated coupling terms could be neglected resulting into two linear independent equations. The resulting amplitudes are
\begin{multline}
\begin{aligned}
	\label{C013CO}
	\displaystyle  C_n= \frac{\sqrt{2\gamma_n}}{i\gamma_n-\Omega_n}B^{\text{in}}_n e^{-i(\varphi_n-\phi_m)},\\
\end{aligned}
\end{multline}
where $\Omega_n = \omega_n-\omega_{pn}$ and phases are chosen to set the coefficents $C_n$ real. 

On the second step, the equations of motion are written for small varying fluctuations $\widetilde{c}_n$, $\widetilde{a}$ and $\widetilde{b}_n$ where only terms of the first order in $\widetilde{c}_n$ and $\widetilde{a}$ are retained.
Denoting $\Gamma_n = i\Omega_n+\gamma_n$, the equations of motion in the Markovian and "slowly varying envelope" approximations:
\begin{multline}
\begin{aligned}
	\label{C016CO}
	\displaystyle  \frac{d}{dt}\widetilde{c}_n = -\Gamma_n \widetilde{c}_n+g_\text{eff}\xi_+ \widetilde{a} C_m e^{-i\Delta_\text{D}t+i\phi_m}\\
	 \displaystyle - i\sqrt{2\gamma_n}\widetilde{b}_n^{\text{in}},\\ 
\end{aligned}
\end{multline}
where $\Delta_\text{D} = \omega_a - \omega_{p1}-\omega_{p2}$ is axion detuning in the downconversion case that can be set to zero. This EOM can be easily solved by transforming the problem into the frequency domain:
\begin{multline}
\begin{aligned}
	\label{C017CO}
	\displaystyle   \widetilde{c}_n^\text{D}[\Omega]=-\frac{g_\text{eff}\xi_+ C_m}{(i\Omega-\Gamma_n)} \widetilde{a}[\Omega]e^{i\phi_m}\\
	\displaystyle - i\frac{\sqrt{2\gamma_n}}{(i\Omega-\Gamma_n)}\widetilde{b}_n^{\text{in}}[\Omega].\\
\end{aligned}
\end{multline}

For the upconversion case, the analysis follows the same steps. Moreover, since on the zeroth order step considering the strong pump case does not include the axion coupling terms, the solution Eq.~(\ref{C013CO}) is valid for this case as well. And, the equation of motion for the small fluctuations are then given as:
\begin{multline}
\begin{aligned}
	\label{C016FT}
	\displaystyle  \frac{d}{dt}\widetilde{c}_n = -\Gamma_n \widetilde{c}_n-g_\text{eff}\xi_- \widetilde{a} C_m e^{-i\Delta_\text{U}t-i\phi_m} \\
	\displaystyle - i\sqrt{2\gamma_n}\widetilde{b}_n^{\text{in}},\\ 
\end{aligned}
\end{multline}
where $\Delta_\text{U} = \omega_a + \omega_{pm}-\omega_{pn}$ is axion detuning in the upconversion case. The solution in the frequency domain can be given as follows:
\begin{multline}
\begin{aligned}
	\label{C017FT}
	\displaystyle   \widetilde{c}_n^\text{U}[\Omega]=\frac{g_\text{eff}\xi_- C_m}{(i\Omega-\Gamma_n)} \widetilde{a}[\Omega]  e^{-i\phi_m}\\
	\displaystyle - i\frac{\sqrt{2\gamma_n}}{(i\Omega-\Gamma_n)}\widetilde{b}_n^{\text{in}}[\Omega].\\
\end{aligned}
\end{multline}

In both upconversion and downconversion cases, Equations (\ref{C017FT}) and (\ref{C017CO}), the mode fluctuations consist of a noise component associated with the input signal $\widetilde{b}[\Omega]$ and signal component $\widetilde{a}[\Omega]$ which is amplified by the strong signal deposited in the other mode $C_m$.

\section{Axion Generated Phase Spectrum in a Dual Mode Cavity} 
\label{A1}

In this Appendix we consider a case when an axion signal is searched in the Fourier spectrum of phase noise of the double mode cavity. In this case, we relax the requirement of the axion frequency to be exactly the sum or difference of resonant frequencies of two modes. Instead, this property is only approximately held true where the detuning constitute the Fourier frequency. In this case, $\omega_a = \omega_1\pm\omega_2+\Omega$ where $\Omega \ll \omega_1$. Under this condition, the axion amplitude appear as a slowly varying parameter in the EOMs. The objective of this derivation is to find a transfer function from this small and slow time-varying quantity to amplitudes and phase fluctuations of the output signals of the cavity. The transfer function is derived by transferring the equations of motion for the complex amplitudes to the equations for real signal phases and magnitude and linearizing them around a steady state point.  


The starting point of this derivation is the equations of motion for the complex amplitudes $C_1$ and $C_2$ of the two modes in the upconversion case. Additionally, we introduce two pump signals $B_1$ and $B_2$, so the equations of motion can be written as follows:
\begin{multline}
\begin{aligned}
	\label{R001VB}
	\displaystyle  \frac{d}{d\tau}C_1 = (-\gamma_1-i\Delta_1(\tau)) C_1-g_\text{eff}\xi_- A(\tau) C_2 + B_1(\tau),\\
	\displaystyle  \frac{d}{d\tau}C_2 = (-\gamma_2-i\Delta_2(\tau)) C_2+g_\text{eff}\xi_- A^\ast(\tau) C_1 + B_2(\tau),\\
\end{aligned}
\end{multline}
where $\tau$ is the "slow time". Time variation of the detuning coefficients $\Delta_i$ accounts for fluctuations of the resonance frequencies of the modes as well as DC detuning of the pump signals. Instead of complex amplitudes, we need to rewrite the equations of motion in terms of real phases and amplitudes: $C_i(\tau) = x_i(\tau)\exp(-j\varphi_i(\tau))$ and  $B_i(\tau) = y_i(\tau)\exp(-j\theta_i(\tau))$. Here $y_i$ and $\theta_i(\tau)$  represents amplitude and phase fluctuations of the pump signals. The EOMs in the real amplitude-phase representations are written as follows:
\begin{multline}
\begin{aligned}
	\label{R002VB}
	\displaystyle \frac{d}{d\tau}x_1 = -\gamma_1 x_1 -  x_2Q_c(\tau) + y_1\cos(\theta_1-\varphi_1),\\
	\displaystyle \frac{d}{d\tau}\varphi_1 = \Delta_1(\tau) -  \frac{x_2}{x_1}Q_s(\tau) + \frac{y_1}{x_1}\sin(\theta_1-\varphi_1),\\
	\displaystyle \frac{d}{d\tau}x_2 = -\gamma_2 x_2 +  x_1Q_c(\tau) + y_2\cos(\theta_2-\varphi_2),\\
	\displaystyle \frac{d}{d\tau}\varphi_2 = \Delta_2(\tau)  -  \frac{x_1}{x_2}Q_s(\tau) + \frac{y_2}{x_2}\sin(\theta_2-\varphi_2),
\end{aligned}
\end{multline}
where $Q_c$ and $Q_s$ are two quadrature of the axion signal defined as follows:
\begin{multline}
\begin{aligned}
	\label{R002VBd}
	\displaystyle Q_c(\tau) = g_\text{eff}\xi_-|A|\cos(\theta_a(\tau)+\varphi_2-\varphi_1),\\
	\displaystyle Q_s(\tau) = g_\text{eff}\xi_-|A|\sin(\theta_a(\tau)+\varphi_2-\varphi_1),
\end{aligned}
\end{multline}
and $\theta_a$ is the axion phase. 

These are nonlinear equations with varying coefficients that cannot be solved exactly. Instead, we linearise them for small signal fluctuations around large steady state amplitudes. Here we assume that both axion signal and input phase and amplitude fluctuations are very small. Thus, we can split each variable and time varying parameter into two components strong steady and small fluctuating: $x_i=\overline{x}_i + \widetilde{x}_i$, $y_i=\overline{y}_i + \widetilde{y}_i$, $\varphi_i=\overline{\varphi}_i + \widetilde{\varphi}_i$ and $\theta_i=\overline{\theta}_i + \widetilde{\theta}_i$. As for the axion signal, it has only the small fluctuating part as described in the introduction to this Appendix. The steady state equations for the constant part are:
\begin{multline}
\begin{aligned}
	\label{R003VB}
	\displaystyle \gamma_1 \overline{x}_1  = \overline{y}_1\cos(\overline\theta_1-\overline\varphi_1),\\
	\displaystyle 0 =\overline\Delta_1  \overline{x_1}  + \overline{y_1}\sin(\overline{\theta}_1-\overline\varphi_1),\\
	\displaystyle \gamma_2 \overline{x}_2  = \overline{y}_2\cos(\overline\theta_2-\overline\varphi_2),\\
	\displaystyle 0 = \overline\Delta_2 \overline{x}_2  + \overline{y_2}\sin(\overline\theta_2-\overline\varphi_2),
\end{aligned}
\end{multline}
that are two independent sets of two equations. Solving these equations give the steady state point $(\overline{y}_i,\overline\varphi_i)$ in terms of input signal parameters $(\overline{y}_i, \overline{\theta}_i)$ and detunings $\Delta_i$:
\begin{multline}
\begin{aligned}
	\label{R003VB}
	\displaystyle \overline{x}_i = \frac{\overline{y}_i }{\sqrt{\gamma_i^2+\Delta_i^2}},
	\displaystyle \overline\varphi_i = \overline\theta_i - \arctan{\frac{\overline{\Delta}_i}{\gamma_i}}.
\end{aligned}
\end{multline}

The linearised EOMs for the small fluctuations are
\begin{multline}
\begin{aligned}
	\label{R004VB}
	\displaystyle \frac{d}{d\tau}\widetilde{x}_1 = -\gamma_1 \widetilde{x}_1 - \overline{x}_2 Q_c(\tau) + \\
	\widetilde{y}_1\cos(\overline\theta_1-\overline\varphi_1)-\overline{y}_1\sin(\overline\theta_1-\overline\varphi_1)(\widetilde\theta_1-\widetilde\varphi_1),\\
	\overline{x}_1\displaystyle \frac{d}{d\tau}\widetilde\varphi_1 =  \overline{x}_1\widetilde\Delta_1  + \widetilde{x}_1\overline\Delta_1  -  \overline{x}_2Q_s\\ + \widetilde{y}_1\sin(\overline\theta_1-\overline\varphi_1) + \overline{y}_1\cos(\overline\theta_1-\overline\varphi_1)(\widetilde\theta_1-\widetilde\varphi_1),\\
	\displaystyle \frac{d}{d\tau}\widetilde{x}_2 = -\gamma_2 \widetilde{x}_2 + \overline{x}_1Q_c(\tau) +\\ \widetilde{y}_2\cos(\overline\theta_2-\overline\varphi_2)-\overline{y}_2\sin(\overline\theta_2-\overline\varphi_2)(\widetilde\theta_2-\widetilde\varphi_2),\\
	\displaystyle \overline{x}_2\frac{d}{d\tau}\widetilde{\varphi}_2 =  \overline{x}_2\widetilde\Delta_2  +\widetilde{x}_2\overline\Delta_2 - \overline{x}_1 Q_s(\tau) \\ + \widetilde{y}_2\sin(\overline\theta_2-\overline\varphi_2) + \overline{y}_2\cos(\overline\theta_2-\overline\varphi_2)(\widetilde\theta_2-\widetilde\varphi_2).
\end{aligned}
\end{multline}
that can be simply written in the matrix form as
\begin{multline}
\begin{aligned}
	\label{R005VB}
	\displaystyle \frac{d}{d\tau}\mathbb{X} = \mathbb{H}\mathbb{X} + \mathbb{A} + \mathbb{N},
\end{aligned}
\end{multline}
where $\mathbb{X} = [\widetilde{x}_1,\widetilde\varphi_1,\widetilde{x}_2,\widetilde\varphi_2]^T$ is a vector of state variables, $\mathbb{A} = [-\overline{x}_2Q_c(\tau),-\overline{x}_2Q_s(\tau)/\overline{x}_1,\overline{x}_1Q_c(\tau),-\overline{x}_1Q_s(\tau)/\overline{x}_1]^T$ is a vector of axion signal, $\mathbb{N}$ is a vector of technical fluctuations coming from the pump signal and internal mode instabilities. The system matrix is  
\begin{multline}
\begin{aligned}
	\label{R005VBs}
	\displaystyle \mathbb{H} = \begin{pmatrix}
 -\gamma_1 & -\overline{\Delta}_1{\overline{x}_1} & 0 & 0 \\
  \frac{\overline{\Delta}_1}{\overline{x}_1} & -\gamma_1 & 0 & 0 \\
  0  & 0  & -\gamma_2 & -\overline{\Delta}_2{\overline{x}_2}  \\
  0 & 0 & \frac{\overline{\Delta}_2}{\overline{x}_2} & -\gamma_2
 \end{pmatrix}.
\end{aligned}
\end{multline}

It is worth noting that in this linearised approximation, all state variable fluctuations as well as the axion signal are fully uncoupled due to their extreme smallness and filtered by the defined system transfer matrix. Moreover, the two modes are completely uncoupled giving two independent signals. The corresponding solutions for technical fluctuations have been previously analysed and can be found elsewhere\cite{Rubiola:2008aa,Goryachev:2011aa,Gbook}. For this reason, we are interested only in axion-phase relationship that is given as follows: 
\begin{multline}
\begin{aligned}
	\label{R006VB}
	\displaystyle  \widetilde{\varphi}_i[s] = \pm \frac{\overline\Delta_i }{s^2+2\gamma_i s + \overline{\Delta}_i^2+\gamma_i^2} \alpha_{j,i}{Q_c[s]} - \\
	 \frac{s + \gamma_i}{s^2+2\gamma_i s + \overline{\Delta}_i^2+\gamma_i^2} \alpha_{j,i}Q_s[s]
\end{aligned}
\end{multline}
where $s$ is the Laplace variable, $\alpha_{j,i} = \frac{\overline{x}_j}{\overline{x}_i}$ is a ratio of stored amplitudes, $+$ sign attributes to the second mode, and the $-$ is for the first one. $Q_c[s]$ and $Q_s[s]$ are Laplace transforms of the two quadratures of the axion signals. It is important to empathise that these quadratures are defined with respect to the phase difference of the cavity modes $\overline\varphi_1-\overline\varphi_2$. Thus, generally the result is phase dependent and could depend on the instance of time when it starts. In other words, the axion signal provides an absolute time scale for this kind of experiment.

In the case when both cavities are pumped on resonance ($\overline\Delta_i=0$), transfer function (\ref{R006VB}) is reduced to the first order low pass filter:
\begin{multline}
\begin{aligned}
	\label{R007VB}
	\displaystyle  H^{\text{U}}[s] = \frac{\widetilde{\varphi}_i[s]}{Q_s[s]} = 
	 \frac{\alpha_{j,i}}{s+\gamma_i} .
\end{aligned}
\end{multline}
It is apparent from this result that the output phase component due to the axion signal is scaled by the ratio of magnitudes in both modes. If one is going to measure $\varphi_1$, it is advantageous to increase the $x_2$ magnitude and keep $x_1$ as small as possible. On the other hand when $ x_1\rightarrow 0$, the carrier signal become undetectable. 

To derive the sensitivity of the setup over a range of Fourier frequencies, we compare result (\ref{R007VB}) to the transfer function for technical phase fluctuations. Without lack of generality, we assume that technical fluctuations are dominated by external fluctuations $\widetilde\theta_i[s]$ whose transfer function in the phase space is $\frac{\gamma_i}{s+\gamma_i}$\cite{Goryachev:2011aa,Gbook}. It can be shown that the internal cavity fluctuations lead to a similar relation. A ratio of the axion signal and phase noise fluctuations at the output of the cavity gives a constant signal-to-noise ratio:
\begin{multline}
\begin{aligned}
	\label{R008VB}
	\displaystyle  \text{SNR}_{i}^\text{U} = {g_\text{eff}\xi_- \alpha_{j,i}} \frac{|A|}{\gamma_i\widetilde{\theta}_i}.
\end{aligned}
\end{multline}
If the cavity is pumped on resonances, the same result is obtained for the case of limits due to external signal fluctuations.



For the downconversion case, the equations of motion in terms of real phases and amplitudes are
\begin{multline}
\begin{aligned}
	\label{R102VB}
	\displaystyle \frac{d}{d\tau}x_1 = -\gamma_1 x_1 +  \overline{x}_2 R_c(\tau) + y_1\cos(\theta_1-\varphi_1),\\
	\displaystyle \frac{d}{d\tau}\varphi_1 = \Delta_1 +  \frac{\overline{x}_2}{\overline{x}_1} R_s(\tau) + \frac{y_1}{x_1}\sin(\theta_1-\varphi_1),\\
	\displaystyle \frac{d}{d\tau}x_2 = -\gamma_2 x_2 +  \overline{x}_1 R_c(\tau) + y_2\cos(\theta_2-\varphi_2),\\
	\displaystyle \frac{d}{d\tau}\varphi_2 = \Delta_2  +  \frac{\overline{x}_1}{\overline{x}_2} R_s(\tau) + \frac{y_2}{x_2}\sin(\theta_2-\varphi_2).
\end{aligned}
\end{multline}
where the axion signal quadratures are
\begin{multline}
\begin{aligned}
	\label{R102VBd}
	\displaystyle R_c(\tau) = g_\text{eff}\xi_+|A|\cos(\theta_a(\tau)-\varphi_2-\varphi_1),\\
	\displaystyle R_s(\tau) = g_\text{eff}\xi_+|A|\sin(\theta_a(\tau)-\varphi_2-\varphi_1).
\end{aligned}
\end{multline}

The solution of the corresponding steady state equation is the same as in the upconversion case described above. Dynamics of the small magnitude-phase fluctuations can be described by the same matrix equation (\ref{R005VB}) and system matrix (\ref{R005VBs}) with a different axion input vector: $\mathbb{A} = [\overline{x}_2R_c(\tau),\overline{x}_2R_s(\tau)/\overline{x}_1,\overline{x}_1R_c(\tau),\overline{x}_1R_s(\tau)/\overline{x}_1]^T$.
The corresponding axion-phase relationship appears as follows
\begin{multline}
\begin{aligned}
	\label{R103VB}
	\displaystyle  \widetilde{\varphi}_i[s] =  \frac{\overline\Delta_i }{s^2+2\gamma_i s + \overline{\Delta}_i^2+\gamma_i^2} \alpha_{j,i}{R_c[s]} + \\
	 \frac{s + \gamma_i}{s^2+2\gamma_i s + \overline{\Delta}_i^2+\gamma_i^2} \alpha_{j,i}R_s[s],
\end{aligned}
\end{multline}
which is different from the upconversion solution (\ref{R006VB}) only in the signs of each term as well as definition of the axion quadratures. In these equations $R_c[s]$ and $R_s[s]$ are Laplace transform of the axion signal defined as in (\ref{R102VBd}). Here the axion signal quadratures are defined with respect to the sum of the phases of signals in both modes $\overline\varphi_1+\overline\varphi_2$. Thus like in the upconversion case, experimental results could be phase sensitive. In the case of two cavities pumped on resonance, the result is also reduced to the first order transfer low pass filter function:
\begin{multline}
\begin{aligned}
	\label{R104VB}
	\displaystyle  H^{\text{D}}[s] = \frac{\widetilde{\varphi}_i[s]}{R_s[s]} =  
	 \frac{\alpha_{j,i}}{s+\gamma_i}. 
\end{aligned}
\end{multline}
This result is identical to that of the upconversion case ($\ref{R007VB}$) up to the definition of the axion quadratures and the geometry factor. The corresponding signal-to-noise ratio is 
\begin{multline}
\begin{aligned}
	\label{R105VB}
	\displaystyle  \text{SNR}_{i}^\text{D} = {g_\text{eff}\xi_+\alpha_{j,i}}  \frac{|A|}{\gamma_i\widetilde{\theta}_i}.
\end{aligned}
\end{multline}

\section{Axion Generated Phase Spectrum in a Dual Loop Oscillator}
\label{oscil}

A loop oscillator is a frequency selective positive feedback system in which two conditions of existence of sustained oscillations are fulfilled: small signal gain across the loop is greater than 1 and open loop phase shift is integer multiple of $2\pi$. These conditions make the small signal solution for the oscillator divergent, although due to amplitude limiting nonlinearity the self sustained signals gets saturated as some value of circulating power. To analyse phase-amplitude fluctuations in such a system, in addition to cavity parameters, one needs to make assumptions on the type of nonlinearity, amplifier gain and phase shift as well as amplifier-cavity couplings. In this case, it is possible to apply the same type of analysis as described in Appendix~\ref{A1}. On the other hand, one may assume a certain level of circulating power, e.g in the form of magnitudes in the cavity modes $\overline{x}_1$ and $\overline{x}_2$, and analyse the system in the small signal regime. In this regime, an ideal amplifier has a unity transfer function in the phase space. The resulting system for the $i$th mode is demonstrated in Fig.~\ref{closedloop}. The phase fluctuations at the output of the oscillator are found as a sum of contributions from the technical noise $\widetilde\theta_i$ and axion signal $R_s$ (or $Q_s$):
\begin{multline}
\begin{aligned}
	\label{R104VB}
	\displaystyle  \widetilde\varphi_i^\text{D} = 
	 \displaystyle\Big[1+\frac{\gamma_i}{s}\Big]\Big(\widetilde\theta_i + \frac{\alpha_{j,i}}{s+\gamma_i}R_s\Big),\\
	 \widetilde\varphi_i^\text{U} = 
	 \displaystyle\Big[1+\frac{\gamma_i}{s}\Big]\Big(\widetilde\theta_i + \frac{\alpha_{j,i}}{s+\gamma_i}Q_s\Big),\\
\end{aligned}
\end{multline}
where the first term constitutes the Leeson effect\cite{leeson} for the technical phase fluctuations in the loop and the second term represents the axion induced phase fluctuations. 

\begin{figure}
\includegraphics[width=0.8\columnwidth]{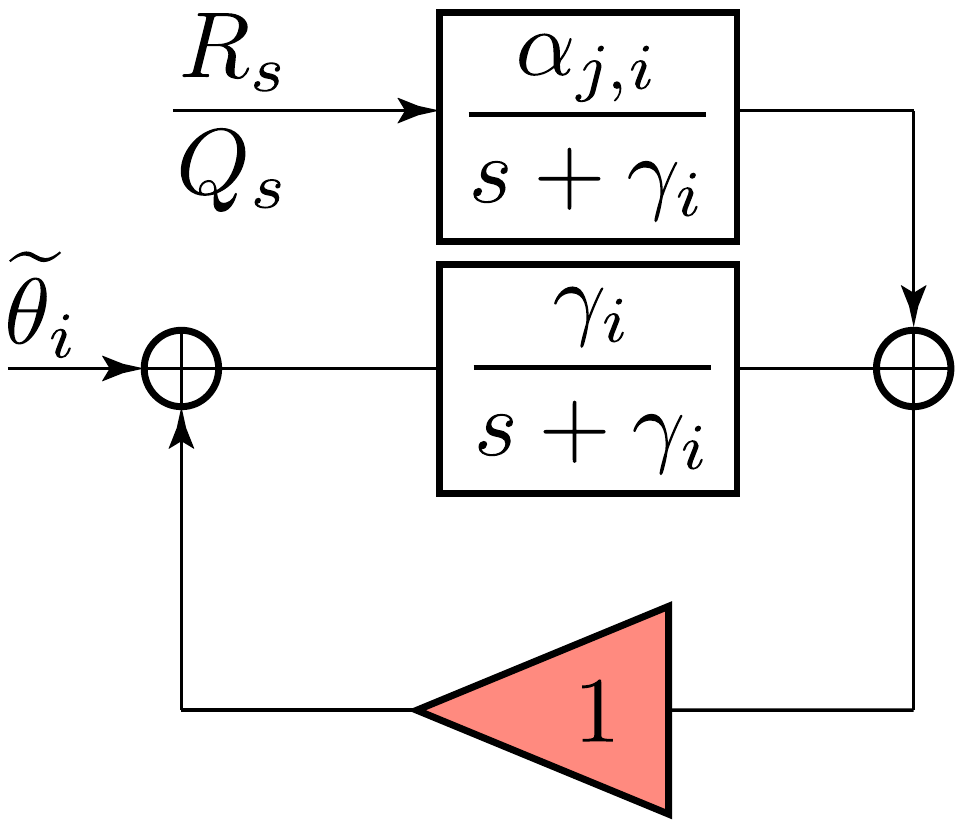}
\caption{Schematic representation of the axion to phase conversion in a feedback oscillator in the phase space. }
\label{closedloop}
\end{figure}


\end{document}